\documentclass[
    ,final            
  ,numberedheadings 
  ]
  {aipproc}

\layoutstyle{6x9}

\begin{document}

\title{Spin fluctuations in the magnetically ordered phase of frustrated 
pyrochlore systems}

\classification{31.30.Gs; 75.50.Lk; 76.30.Kg; 76.80.+y, 76.75.+i}
\keywords      {4f pyrochlore systems, spin fluctuations, geometrically
frustrated magnetism}

\author{P. Bonville}{
  address={CEA, CE Saclay, DSM/Service de Physique de l'Etat Condens\'e,
91191 Gif-sur-Yvette, France}
}

\author{I. Mirebeau}{
  address={CEA, CE Saclay, Laboratoire L\'eon Brillouin, 91191 Gif-sur-Yvette,
France}
}

\author{J.-P. Sanchez}{
  address={CEA, CE Grenoble, DSM/Service de Physique Statistique, Magnétisme
et Supraconductivit\'e, 38054 Grenoble, France}
}

\begin{abstract}
Two geometrically frustrated pyrochlore stannates, undergoing long range
magnetic order below 1\,K, were investigated at very low temperature.
Anomalies in the behaviour of hyperfine quantities are found, by $^{155}$Gd 
M\"ossbauer spectroscopy in Gd$_2$Sn$_2$O$_7$ and by 
low temperature specific heat measurements in Tb$_2$Sn$_2$O$_7$. They are 
interpreted in terms of fluctuations of the correlated Gd or Tb spins, using a
model two-level system (the nuclear spins) submitted to a randomly fluctuating
(hyperfine) field.
\end{abstract}

\maketitle


\section{Introduction} \label{intro}
Geometrical frustration in magnetic systems occurs, for instance, in 
crystallographically ordered
\textit{kagom\'e} or pyrochlore lattices, where the building unit is 
resp. a triangle and a tetrahedron, and where the units are loosely bound, in
general sharing an apex only (see Ref.\cite{ramirez} for a review). The simple
example of Heisenberg (isotropic) spins on a triangle, coupled by 
antiferromagnetic (AF) exchange, illustrates the impossibility for this 
system to reach a collinear N\'eel type ground state. Frustration also plays
a role in the case of Ising spins, but for ferromagnetic interactions, leading
to the so-called ``spin-ice'' ground state \cite{harris}, as found in the 
pyrochlore Ho$_2$Ti$_2$O$_7$. Geometrical frustration is expected
to prevent the onset of long range magnetic order (LRO) down to the lowest
temperature, but extra interactions like the dipolar coupling in the AF 
Heisenberg case, or a finite anisotropy in the case of spin-ices, can lift
the large degeneracy of the ground state and stabilize a magnetic LRO state.

We focus here on two frustrated pyrochlore stannates undergoing LRO 
below around 1\,K, Gd$_2$Sn$_2$O$_7$ \cite{wills} and Tb$_2$Sn$_2$O$_7$
\cite{mirebeau}. We report on indirect experimental evidence, through the
measurement of hyperfine quantities, that fluctuations
of the correlated spins persist in the LRO phase, at least down to the 0.1\,K
range. Before we present the data, we will develop the model which mimics the
real situation, i.e. a nuclear spin submitted to a randomly fluctuating
electronic (hyperfine) magnetic field \cite{bertin}.    

\section{Two-level system driven by a randomly reversing field} \label{twolv}

We consider a (nuclear) spin 1/2, whose levels are split by a magnetic field, 
and we assume that two
time scales govern the dynamics of the system: a relaxation time $T_1$, which
maintains the equilibrium populations of the two levels, and a time 
$\tau$ associated with fluctuations of the field. At low temperature in the
LRO phase, $T_1$ can be viewed as a magnon driven spin lattice relaxation
time. We wish to calculate the steady state average populations of the two
levels as a function of the ratio $T_1/\tau$, which in turn allows to obtain
two hyperfine
quantities of interest: the effective hyperfine temperature T$_{hf}$ and
the hyperfine (nuclear) specific heat $C_{hf}$. When $\tau \gg T_1$,
i.e. the common case, the hyperfine levels have time to reach thermal 
equilibrium in the interval between two reversals of the field; 
then T$_{hf}$ = T and $C_{hf}=
C_{Sch}$, where $C_{Sch}$ is the regular Schottky anomaly associated with
the hyperfine
splitting $\Delta_{hf}$. Conversely, when $T_1 \gg \tau$, the hyperfine 
levels are 
completely out of equilibrium, the steady state populations of the two levels
are equal, and T$_{hf}$ is infinite. For a finite value of the ratio 
$T_1/\tau$, it turns out that 
the problem is analytically solvable, and the details of the calculation can 
be found in Ref.\cite{bertin}. The steady state population $p_g$ of the ground
level writes:
\begin{equation}
p_g({\rm T}) = \frac{1}{2}\left(1+\frac{1}{1+2\frac{T_1}{\tau}}
\ \tanh\frac{\Delta_{hf}} {k_B{\rm T}}\right),
\end{equation}
the hyperfine temperature being related to $p_g$ by: $p_g({\rm T}) =
\left(1+\exp(-\frac{\Delta_{hf}}{k_B {\rm T}_{hf}({\rm T})})\right)^{-1}$.
Then, when $k_B{\rm T}>2\Delta_{hf}$, one obtains: T$_{hf} \simeq$ 
T(1+2$\frac{T_1}{\tau})$, which means that the out of equilibrium hyperfine 
system is warmer than the lattice, and:
\begin{equation}
C_{hf} = \frac{1}{1+2\frac{T_1}{\tau}}\ C_{Sch}+\alpha \ \Delta_{hf} 
{\rm\frac{d}{dT}}(\frac{T_1}{\tau}), 
\label{cpred}
\end{equation}
where $\alpha$ is a dimensionless coefficient of order unity. Our estimations
show that the second term in this expression is much smaller than the
first one, with the reasonable assumption that $T_1/\tau$ varies slowly with
temperature. Then, the hyperfine specific heat is reduced by a factor
$1+2\frac{T_1}{\tau}$ with respect to
the expected Schottky anomaly, but does not change its shape. This implies
an apparent reduction of the hyperfine field $H_{hf}$ giving rise to the 
nuclear level splittings.

These two remarkable effects, the warming up of the hyperfine levels and the
reduction of the Schottky nuclear anomaly, can be observed when it occurs
that the nuclear relaxation time $T_1$ is of the same magnitude as the 
reversal time $\tau$; the latter corresponds to a spin-flip time since, for
rare earths, the hyperfine field is proportional to the magnetic moment 
to a very good approximation.

\section{The very low temperature $^{155}$Gd M\"ossbauer spectrum in 
Gd$_2$Sn$_2$O$_7$} \label{gdmoss}
  
\begin{figure}
\includegraphics[height=.23\textheight]{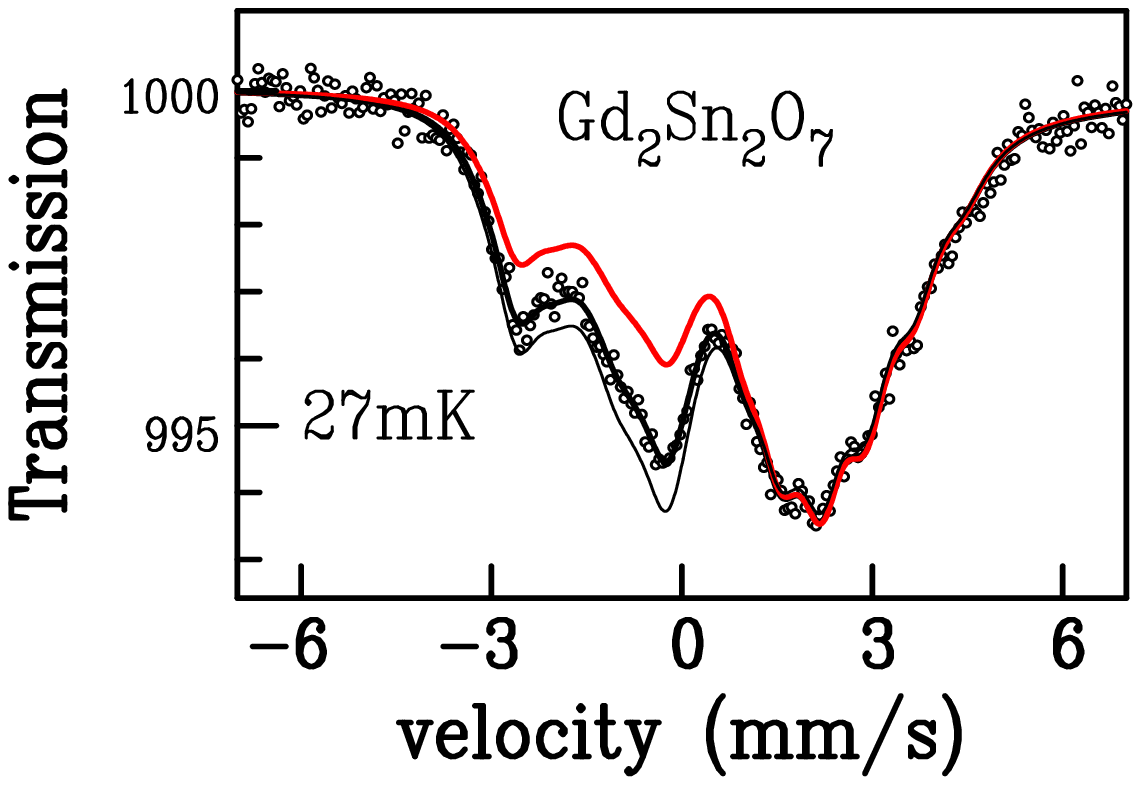}
\hspace{0.3cm}
\includegraphics[height=.27\textheight]{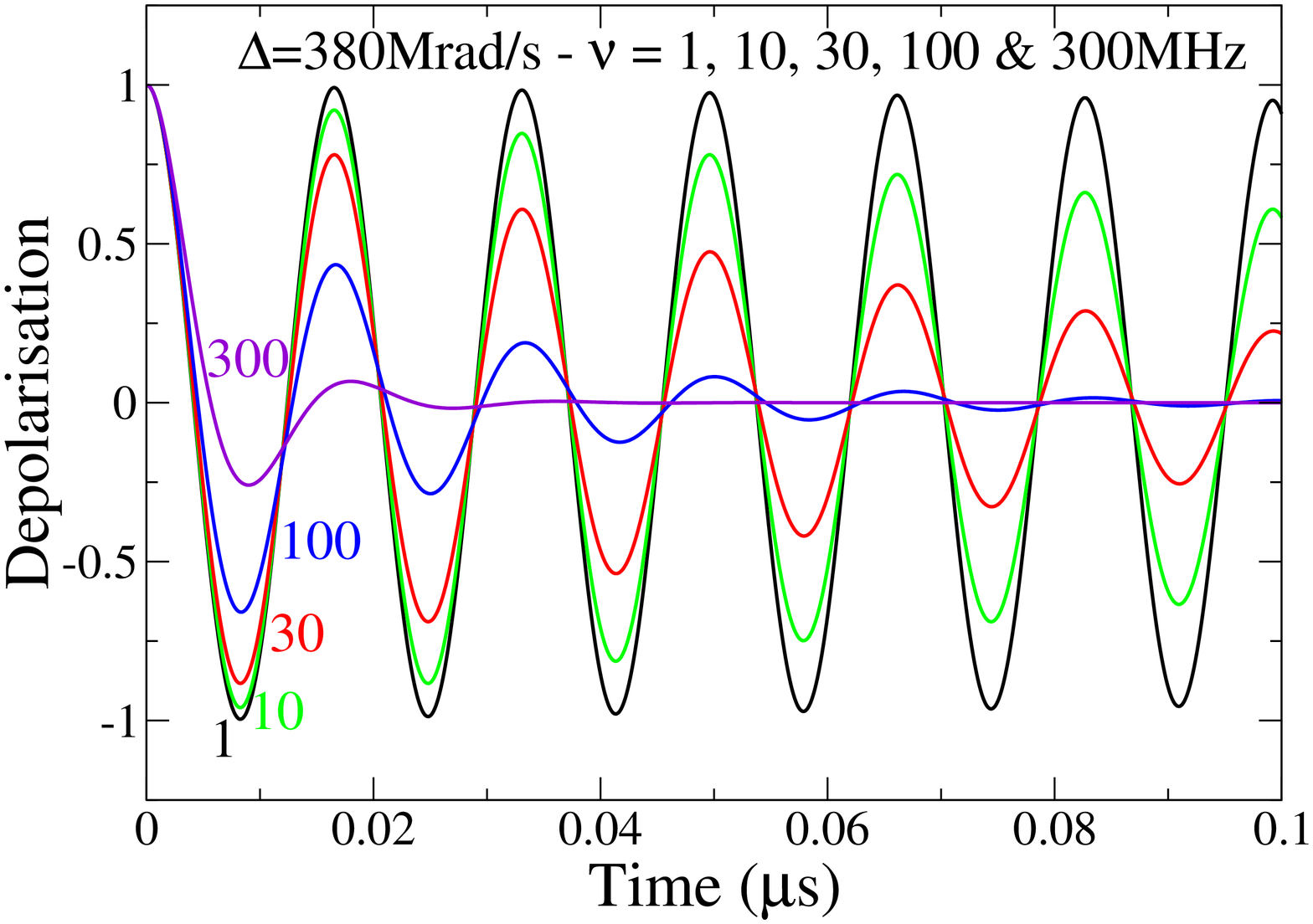}
\caption{\textbf{Left}: $^{155}$Gd M\"ossbauer spectrum at 0.027\,K in 
Gd$_2$Sn$_2$O$_7$; the lines are calculated spectra at different temperatures.
Thick red line: expected spectrum at the lattice temperature 0.027\,K, 
thin black line: expected spectrum at high temperature ($T>0.15$\,K),
thick black line: spectrum with fitted effective hyperfine temperature
$T_{hf}$ = 0.09\,K. \textbf{Right}: Calculated muon depolarisation signal in 
the presence of a fluctuating dipolar field $H_{dip}$, corresponding to 
$\Delta$=380\,Mrad s$^{-1}$, for several fluctuation frequencies $\nu$.}
\label{gdsn}
\end{figure}

Gd$_2$Sn$_2$O$_7$ shows magnetic order below 1\,K and its AF \textbf{k}=0
magnetic structure corresponds to that expected for a Heisenberg pyrochlore
antiferromagnet where the degeneracy is lifted by a sizeable dipolar 
interaction \cite{palmer}. The $^{155}$Gd M\"ossbauer spectrum at 0.027\,K
is shown in Fig.\ref{gdsn} left. It is a LRO spectrum with a
hyperfine field $H_{hf} \simeq 30$\,T, and with a sizeable hyperfine 
quadrupolar interaction. The ground
nuclear spin $I_g=3/2$ is then split into two quasi-doublets separated by 
about 0.015\,K. 
Thus, below 0.1\,K, the line intensities of the M\"ossbauer transitions depend
on temperature through the Boltzmann populations of the hyperfine 
levels, and the absolute temperature can be determined by fitting the spectrum.
It is clear in Fig.\ref{gdsn} left that the hyperfine temperature
(0.09\,K) is higher than the lattice temperature (0.027\,K). According
to section \ref{twolv}, this points to the presence of Gd spin fluctuations,
at 0.027\,K,
with a time scale $\tau$ of the same order as the nuclear spin-lattice time
$T_1$: $T_1/\tau \simeq 1$. Since no relaxational effects are observed in the 
M\"ossbauer spectrum,
this implies that $\tau$ is longer than the hyperfine Larmor period for
$^{155}$Gd $\tau_L \sim 3\times 10^{-8}$\,s. Muon spin resonance ($\mu$SR)
measurements carried out at 0.02\,K in zero field and longitudinal geometry
\cite{bonvi} show an oscillating time decay due to the precession of the 
$\mu^+$ spin in the dipolar field $H_{dip}$ from the Gd moments. The pulsation
associated with the precession is $\Delta = \gamma_\mu H_{dip}$, where 
$\gamma_\mu = 
851.6$\,Mrad s$^{-1}$T$^{-1}$ is the $\mu^+$ gyromagnetic ratio. Actually, 
one observes two pulsations:  $\Delta$=190 and 380\,Mrad s$^{-1}$, 
corresponding probably to two muon stopping sites. 
No damping at short times is observed, and this allows to set a lower limit 
for $\tau$. 

For this purpose, we computed the $\mu^+$ depolarisation $P_z(t)$ in the 
presence of a dipolar field perpendicular to the initial $\mu^+$ polarisation
direction and 
randomly reversing with a frequency $\nu$. We used the stochastic theory
developed in Ref.\cite{datta}, which assumes the field jumps have a stationary
and Markovian character. In the slow relaxation
regime ($\nu < 2\Delta$), one obtains a damped oscillatory behaviour:
\begin{equation}
P_z(t) = \exp(-\frac{\nu}{2}\ t) \left(\cos\delta t+\frac{\nu}{2\delta}
\sin\delta t \right),
\label{depfl}
\end{equation}
where $\delta=\sqrt{\Delta^2-\nu^2/4}$. Setting $\beta=\frac{\nu}{2\Delta}$,
this signal can be viewed as 
a damped cosine function with a reduced pulsation $\Delta' = \Delta \sqrt{1-
\beta^2}$ and a phase shift $\varphi$ such that $\tan \varphi = \beta/\sqrt{1
-\beta^2}$. In the limiting case $\nu = 2\Delta$, one gets: $P_z(t) = \exp(-
\frac{\nu}{2}t)\ (1+\frac{\nu}{2}t)$. 

In the fast relaxation regime ($\nu >
2\Delta$), an expression analogous to (\ref{depfl}) holds, with the 
trigonometric functions
replaced by hyperbolic ones, and with $\delta = \sqrt{\nu^2/4-\Delta^2}$.
In this case, an exponential-like decay is obtained. In the extreme narrowing 
limit, when $\nu \gg 2\Delta$, a true exponential decay occurs: $P_z(t) =
\exp(-\lambda_z t)$, with $\lambda_z = \frac{\Delta^2}{\nu}$. This latter
formula differs by a factor 2 from the usual expression for $\lambda_z$ in
the paramagnetic phase, i.e. in the presence of randomly distributed
fluctuating fields. 

The behaviour of $P_z(t)$ is sketched in Fig.\ref{gdsn} right, for $\Delta=
380$\,Mrad s$^{-1}$ and for fluctuation frequencies ranging from 1 to 
300\,MHz. It can be seen that the upper limit of the fluctuation frequency 
which allows
the full undamped oscillations to be observed amounts to about 10 MHz. The 
lower limit for $\tau$ is then 10$^{-7}$\,s, which is compatible with the
M\"ossbauer data.

For Gd materials, the small hyperfine interaction precludes the observation of
the nuclear Schottky anomaly in the currently attainable temperature range for
specific heat measurements (T$ > 0.05$\,K).
 
\section{Low temperature specific heat in Tb$_2$Sn$_2$O$_7$}

\begin{figure}
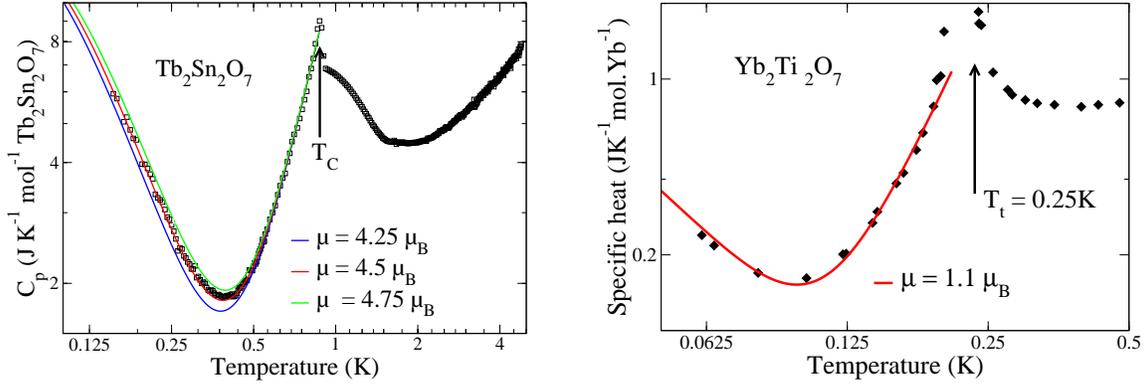

\includegraphics[height=.23\textheight]{cp1n.eps}
\hspace{0.7cm}
\includegraphics[height=.23\textheight]{cp_blote.eps}
\caption{Specific heat in Tb$_2$Sn$_2$O$_7$ (\textbf{left}) and 
Yb$_2$Ti$_2$O$_7$ (\textbf{right}), the latter from Ref.\cite{blote}. The 
solid lines are calculated as the sum of a nuclear Schottky anomaly, with 
several hyperfine field (and hence magnetic moment) values for 
Tb$_2$Sn$_2$O$_7$, and a hyperfine field of 114\,T for Yb$_2$Ti$_2$O$_7$, from
$^{170}$Yb M\"ossbauer measurements \cite{hodges}, and of a magnon term with
thermal dependence resp. $T^3$ and $T^4$.}
\label{tbsn}
\end{figure}

Tb$_2$Sn$_2$O$_7$ orders magnetically below 0.87\,K, according to a 
\textbf{k}=0 ``ordered
spin-ice'' structure \cite{mirebeau}, i.e. the Tb$^{3+}$ magnetic moments lie 
close to the four threefold $<111>$ axes within a tetrahedron in the 
``two in-two out'' configuration, like in a regular spin-ice. 
The local saturated Tb$^{3+}$ moment is found to be 
5.9(1)\,$\mu_B$/Tb. The rare-earth magnetic ($4f$) specific heat (the lattice
contribution is negligible below 5\,K) is shown in Fig.\ref{tbsn} left.  
Apart from the $\lambda$-like anomaly at $T_C$, a nuclear Schottky tail is
observed in the LRO phase below 0.4\,K. It can be accounted for by a 
hyperfine field of 180\,T acting on the $I=3/2$ spin of the isotope 
$^{159}$Tb, the hyperfine quadrupolar contribution being very small. Since
the hyperfine constant for Tb is 40(3)\,T/$\mu_B$ \cite{dunlap}, the 
corresponding magnetic moment is $m$=4.50(3)\,$\mu_B$/Tb \cite{note} 
(see solid 
lines in Fig.\ref{tbsn} left). This value is remarkably smaller than the 
value 5.9\,$\mu_B$
found in the neutron diffraction experiments \cite{mirebeau}, which we 
interpret, in the frame of the model developed in section \ref{twolv}, as
a clue to the presence of spin fluctuations. Indeed, the high temperature tail
of the nuclear Schottky anomaly is proportional to $\Delta_{hf}^2$, 
i.e. to $m^2$. Hence, an overall 
depletion of the specific heat according to expression (\ref{cpred}) leads to
a reduction of the derived moment by the factor $\sqrt{1+2\frac{T_1}{\tau}}$. 
In the present case, this yields: $T_1/\tau \simeq 0.4$. Fluctuations of the
correlated Tb spins in the LRO phase were also inferred from $\mu$SR data
\cite{dalmas,bert}. Contrary to Gd$_2$Sn$_2$O$_7$, no oscillations are
observed, and the time decay of the $\mu^+$ spin depolarisation is
exponential, which corresponds to the extreme narrowing limit
of expression (\ref{depfl}). From the measured low temperature value 
$\lambda_z \simeq 2.3$\,MHz and a value 
$\Delta \sim 100$\,Mrad s$^{-1}$, $\tau$ is estimated at 10$^{-10}$\,s
\cite{dalmas}.

In the parent compound Tb$_2$Ti$_2$O$_7$ which, contrary to Tb$_2$Sn$_2$O$_7$,
does not order down to 0.05\,K, a Schottky upturn in the specific heat
has nevertheless been observed below 0.3\,K\cite{siddha}. It corresponds
approximately to a hyperfine field of 115\,T, i.e. to a Tb moment value of 
2.9\,$\mu_B$, which is also strongly reduced with respect to the crystal
field ground state value of $\simeq 5$\,$\mu_B$ \cite{gingr,mirb}. This is
very likely to be due to moment fluctuations, associated with the strong
short range dynamic spin correlations which persist down to the lowest 
temperature
in the spin-liquid Tb$_2$Ti$_2$O$_7$.

\section{Summary and conclusions} 

Anomalies in the hyperfine quantities were detected in
the LRO phase of some rare-earth based frustrated pyrochlore systems. They 
strongly suggest that spin fluctuations of the correlated moments persist  
in the magnetically orered phase, down to very low temperature, with 
frequencies in the range 10\,MHz - 10\,GHz. The 
mechanism underlying these fluctuations could be a tunneling between 
degenerate spin configurations. However, these observations demand that
the two time scales, the nuclear relaxation time $T_1$ and the spin-flip time
$\tau$, be of the same order of magnitude, which is strongly material 
dependent. For example, no anomaly in the hyperfine temperature is observed in
Gd$_2$Ti$_2$O$_7$ \cite{bertin}, whereas this material is likely to behave
similarly to Gd$_2$Sn$_2$O$_7$. The nuclear Schottky upturn in
Yb$_2$Ti$_2$O$_7$ \cite{blote} corresponds exactly to the hyperfine field
of 114\,T measured by $^{170}$Yb M\"ossbauer spectroscopy \cite{hodges}
(see Fig.\ref{tbsn} right), although spin fluctuations have been evidenced
in the low temperature short range ordered phase of this material 
\cite{hodges}.
 
\begin{theacknowledgments}
We are grateful for useful discussions with P. Dalmas de R\'eotier and
J.A. Hodges.
\end{theacknowledgments}

\bibliographystyle{aipprocl} 

\begin{thebibliography}{99}
\bibitem{note}
A moment value of 3.38$\mu_B$/Tb has been erroneously stated in 
Ref.\cite{mirebeau}.
\end{thebibliography}

\end{document}